# A New Trusted Third-Party Billing for Mobile Networks

Supriya Chakraborty[1] and Nabendu Chaki[2]

[1]Department of Computer Application, JIS College of engineering, West Bengal, India
`supriya.k6@gmail.com`
[2]Department of Computer Science & Engineering, University of Calcutta, India
`nabendu@ieee.org`

## ABSTRACT

*The advancement of technology facilitates explosive growth of mobile usage in the last decade. Numerous applications have been developed to support its usage. However, gap in technology exists in obtaining correct and trusted values for evaluation indexes of the precise amount of usage. The claims of loss in revenue by the service providers could be more due to unexpected behaviour of the hardware. A similar mistrust is often observed in the users of the services. A trustworthy subscription scheme is in demand for consumers whereas revenue needs to be assured of the service providers. Multiple Authorizations by Multiple Owners (MAMO) has already been introduced as a technology to build trust in the third party billing system. In this paper, MAMO is extended to ensure trustworthiness of the parameters for subscription. Along with call transaction data are reconciled to assure the proper revenue generation.*

## KEYWORDS

*Trusted Billing Amount, Synchronized Billing scheme, Revenue assurance, Customer Trust*

## 1. INTRODUCTION

Many mobile subscribers are not contentment on subscription adjustment both in prepaid and post-paid connections from the date of inception of its usage. Printed bills with call transaction details are available from the service provider's end; services of customer care support are successfully practised all over the world. But *correctness*, *accuracy* of parameters that are provided by the service providers is in doubt from the very early days of its usage. A protection scheme to ensure the trustworthy of the parameters is in demand to meet consumer satisfaction. In other side service providers need to be assured with the proper revenue generation.

Complain against subscription amount, duration of call, validity of the scheme, and value added services (game, ring tone, caller tune, song, movie etc.) have been logged into the service providers end in the recent years. Alteration of parameters of call transactions data may be happened intentionally or unintentionally. Sometimes distortions of parameters are occurred due to unexpected performance of the system too. Customer satisfaction is already in the judiciary in many of the major countries but still technological gap exist neither to prove claims of consumers nor service providers. Consumers are forced to ignore challenges in the hazard full fast moving life. The distrusted billing system is the primary concern of both sides. Customer must be assured on the accuracy and correctness of all the parameters of call transactions data that are guided by the proven technology. To ensure trustworthy of billing system, a scheme namely MAMO (Multiple Authorizations by Multiple Owners) was proposed in [1]. Along with a synchronized third party billing scheme was discussed.

Service providers also need to be assured on the revenue generation. Their experience in billing is incurred with reduced revenue amount than the number of actual calls made throughout the cycle. Hardware failure, synchronization, overflow etc may be among the reasons for the reduced revenue generation. But a systematic treatment is required to solve this problem.

Revenue assurance with MAMO is proposed in this work. The synchronized billing system is extended to accommodate revenue assurance of the service provider. However trust of billing amount is ensured by the MAMO itself. The extended scheme is contributed with the followings:

- *Service providers are scaled on the top line of revenue generation.* This scheme may correct 5% revenue generation of one service provider but it may ends up with the higher percentage of correction in another organization. Overall, service providers are scaled in no reduced billing than the actual.

- *Trustworthiness of the billing system is ensured..* Important parameters of subscription are protected from alternation both intentionally and unintentionally.

This paper is organized as follows. It commence with the brief summary of the work, following the introduction section. Then related works are summarized just immediate below this paragraph. In the next subsection MAMO is presented with different type of authorizations. Then formulation of Extended MAMO is presented with schematic diagram of the billing scheme in successive sections. In the next section implementation of MAMO is done. Implementation comprises of software requirements, following the result section with tables and graphs. Finally conclusion is drawn of this work following acknowledgment and references. This paper ends with the short biography of authors.

## 2. RELATED WORK

Parameters of subscription, formulation and its computation of telecommunication services need to be pondered for many reasons. The backbone of the telecommunication services are built by the Government fund in most of the countries. This backbone is used in agreement basis and services are provided to the end users by big multinationals. The ratio between revenue generation during a fiscal year and agreement amount for that fiscal between government-multinationals is very high. The percentage of profit of multinationals also includes superfluous subscription computation in telecommunication and their value added services. Thus parameters of subscription need to be reanalysed for balancing between percentage of profit of service providers and current financial burden of customers. At least, consumers are expected to be secured by a trustworthy system in their ends. Few frequently used security techniques are briefed in the following paragraphs.

The performance of loose and high coupling was evaluated to build hybrid network for 3G by [2]. Interestingly low response time, jitter and end to end delay was experienced with the loose coupling that definitely effects *billing*. The *video call billing* was proposed using an embedded watermarking technique by [3]. Each video sequence is transmitted by the embedded watermark to trace the network noise which is evaluated at the recipient end. Learning of new services in 3G environments has been proposed to be quickening by the usage of AJAX in [4]. A *third party billing* system has been proposed by [5]. Context has been enveloped to compute the *real time billing*. A content based billing was proposed by the least common multiple capacity algorithm in 3G environment [6]. The performance result outperforms its contemporary. A different aspect of multi grade services was analyzed in IP multimedia subsystem to provide *flexible charging scheme with QoS provisioning* in [7]. Watermarking, a means of hiding copyright data within images, are becoming necessary components of commercial multimedia applications that are subject to illegal use [8]. Internet is overwhelmed by the digital assets like image, audio etc. In [8], a fragile watermarking scheme has been proposed that detects and locates even minor tampering applied to the image with full recovery of original work. Steganography [9] covers data within audio or video file.

A radically different approach, cantered on natively representing text in fully-fledged databases, and incorporating all necessary collaboration support is presented in [10]. The concept and architecture of a pervasive document editing including management system has been discussed in [11]. Database techniques and real-time updating are used to do collaborative work on multiple devices. COSARR combines a shared word-processor, chat-boxes and private access to internal and external information resources to facilitate collaborative distance writing [12].

Other related important works include classification of text blocks of bibliographic items which compose understanding thesaurus [13]. This method includes incompletely recognized text distantly and utilizes it for reference. It improves the efficiency of digital library. Internet based collaborative writing tool with comparison results are shown in [14]. The Xaxis proposes a framework for web-based collaborative application in an Air Force dealing with dynamic mission re-planning [15]. A developing platform is proposed for collaborative e-commerce on which awareness is the key for users to learn all sorts of information about people and system in this environment [16]. It presents the analysis of traits of awareness information in collaborative e-commerce environment. An analysis method based on rules for software *trustworthiness* is proposed in [17]. It mainly focused on the trustworthiness of software component and lifecycle process. The trustworthy resource extraction rules, analyzing rules and synthesis rules was discussed. However, multiple authorizations in different parts of single document by multiple users with assurance of revenue have not been discussed in the literature as per our investigation till date.

## 2.1. MAMO: Multiple Authorizations by Multiple Owners

A trusted third party billing framework has been proposed based on securing the key parameters for billing using a novel, manifold authorization technique called Multiple Authorizations by Multiple Owners (MAMO). Each of the multiple messages is authorized by the author of the message.

Data would be acquired from the base station as well as mobile handset by the third party billing section. Appropriate authorization of the data will be set priory by both the service provider (base station) and handset. Such authorization will remain enforced until and unless the author (owner) itself changes it.

Message from the base station is considered as primary source for billing. Authorized data would reach via a message from the base station and handset too. Associated massages will be reconciled at the third party billing section.

Different types of authorizations of MAMO are defined as below:

- *Read Only*: The segment cannot be modified by any users.

- *Add Beginning*: Text may be added only at the begging of the segment but nowhere else by any users. Existing text of the segment cannot be modified.

- *Add End*: Text may be added only at the end of the segment but nowhere else by any users. Again existing text of the segment cannot be modified.

- *Add without Altered*: Text may be added anywhere of the segment but existing text cannot be modified by any users.

- *Add with Altered*: Existing text may be modified as well as new text could be added by the users.

Now the challenge is how to implement these authorizations and along with its ownership? The following segment is elaborately discussed with these issues.

The proposed Multiple Authorizations by Multiple Owners (MAMO) methodology is described with a formal representation using the help of theory of automata. Consider the following grammar :

G = (Vn, $\sum$, P, S), where

Vn is a finite, nonempty set whose elements are called variables

$\sum$ is the set of alphabet | set of punctuation symbol | set of integers | $\phi$, ($\phi$ implies Null)

P is a production rule that impose the authorization by the owner of the segment, and S is called the start symbol of Vn.

In this context, the start symbol is implied as section with any granularities (a word, or line or paragraph, or entire document) on which authorization mode is embodied by owner. The authorization mode is checked before accessing the section. If imposed authorization mode is satisfied then user can only access the document. Access privileges are according to the authorization mode for example if owner specifies Add Beginning on the fragment, user can only add text before the fragment but nowhere else. The authorization mode is formulated by a set of production rules set by the owner. Now rules for different authorization mode are specified below:

*Read only* mode is formulated by the following rules:

$S \rightarrow S$ …i

*Add Beginning* mode is formulated by the following rules:

$S \rightarrow S_0 S$ …ii

$S_o \rightarrow wS_o \mid \phi$ …iii

where w is any string that in $\sum$* and $\phi$ implies Null.

*Add End* mode is formulated by the following rules:

$S \rightarrow SS_0$ ...iv

$S_o \rightarrow wS_0 \mid \phi$ …v

*Add without alter* mode is formulated by the following rules:

$S \rightarrow S_1 S_2 S_3 ........ S_n$ vi

$S \rightarrow S_0 S \mid SS_0 \mid S_1 S_0 S_2 ….S_n \mid S_1 S_2 S_0 S_3 ….S_n \mid ….\mid S_1 S_2 S_3 …S_0 S_n,$ …vii

$S_o \rightarrow wS_o \mid S_o w \mid \phi$ …viii

*Add with alter* mode is formulated by the following rules:

$S \rightarrow S_1 S_2 S_3 ……..S_n$ …ix

$S_i \rightarrow \phi$ ; where $S_i$ denotes any of $S_1, S_2, … , S_n$ (x)

$S \rightarrow S_0 S \mid SS_0 \mid S_1 S_0 S_2 ….S_n \mid S_1 S_2 S_0 S_3 ….S_n \mid ….\mid S_1 S_2 S_3 …S_0 S_n,$ …xi

$S_0 \rightarrow wS_0 \mid S_0 w \mid \phi$ …xii

The experimental result would be shown in section 3.4.

## 3. EXTENDED MAMO

### 3.1. Abstract View of Workflow of Call Transactions to Billing Section

An abstract interface to the third party billing system on the workflow of call transactions are depicted in figure 1. Data of only depicted devices are elicited by billing section. Few raw data are also received that are processed as per the requirement by the billing section. A few data fields are enlisted in Table 2 in which raw data fields are boldly formatted.

In figure 1, mobile call is first transferred to the switch. Then switch request to know the validity of the customer to the *Home Location Register* (HLR). If the response gets satisfied then subsequent request is forwarded to the Intelligent Network (IN). The billing scheme in which the customer is currently enrolled is already stored into IN. Finally the call is routed to the destination from the switch and all important parameters are stored regarding call transaction into the IN for billing or balance adjustment of the prepaid connection. In the rest of the paper, billing always refers to the balance adjustment of the prepaid connection.

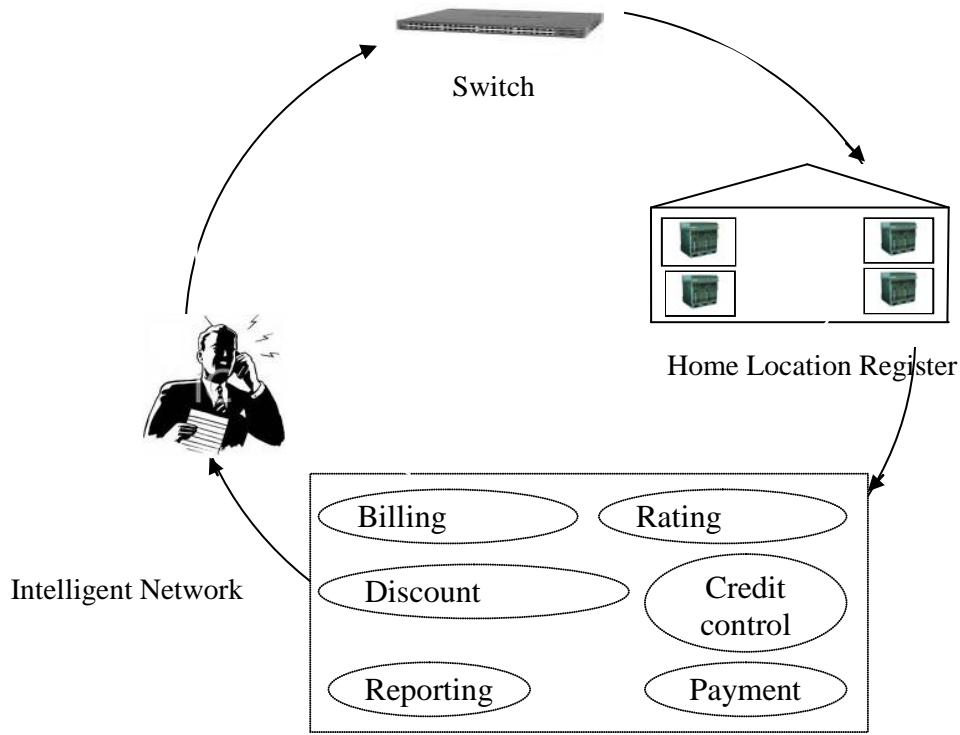

Figure 1: Abstract View of Billing System to Third Party

It has been found that call transactions details of IN deferred from the call details taken from the switch. Any unexpected behaviour of hardware devices is the reason for this reduced call transaction details. The reduced number of call transactions within IN from switch makes the revenue lower than actual of service providers. Our proposed work reconciles call transaction data of IN with the data taken from the switch to ensure revenue of the service provider using MAMO. Activities of workflow in subscriber's handset, IN and switch are depicted in figure 2. The complete billing scheme is shown in figure 3.

### 3.2. Billing Scheme with Extended MAMO

The workflow is already described in section 3.1. In this section the billing scheme is elaborately discussed. The billing section will receive data from three different sources the subscribers hand set, IN and switch. The preparation of messages enveloped by MAMO in these three sources is depicted in figure 2.

Each Call transaction details are stored into IN. Each call detail is covered with MAMO to ensure trustworthy of customers. Then a chunk of call transactions are sent to the billing section. Exactly similarly call transaction details are collected from the subscribers hand set by the billing section. Before acquiring data, MAMO is also applied to the data of the subscribers.

An associated device (AD) with the switch is assumed to ensure revenue generation that need to be plugged in. A software implementation in the managed switch is also envisioned. AD switch also apply MAMO to each call transaction, then buffer authorized data to a limited storage unit. The amount of data could be buffered is implementation specific. Here a unit say x is assumed for simplicity; where x is any positive real number. Buffered data of amount x unit is restored into non-volatile memory. Restoration is done within a time schedule. In that time schedule n number of times x amount of buffered data are restored. That time schedule has a unique number that is generated from the system. The formulation is done as below:

$$T_1 = \{c_1, c_2, c_3, ........................, cn\}$$
$$T_2 = \{cn_1, cn_2, cn_3, .................. cn_n\}$$
$$....................$$
$$T_n = (cn_n.n_1, cn_n.n_2, cn_n.n_3, ................cn_n....n_n) \quad \text{-----------------------(xiii)}$$

Where $T_i$ refers to the unique id of each time schedule in which call transaction details are restored into non-volatile memory storage. Call transaction details are ordered based on the unique call id's ($c_i$).

The AD switch will start probing after completion of each schedule. When it finds a low traffic, restored data is sent to the billing section. The billing section will receive the data ($T_i$) and stored into the storage unit for further processing.

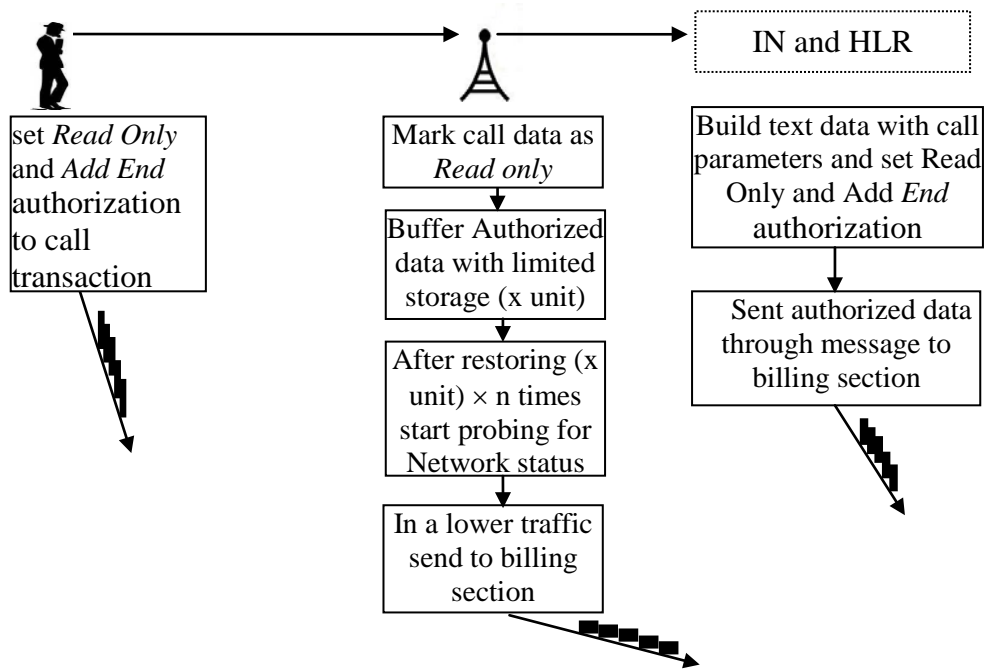

Figure 2. Preparation of Messages in Trust Management

The billing section has two wings separated by the horizontal line in figure 3. Data from IN and subscribers hand set are reconciled by the top wing and putted into the storage unit.

In the top wing, first the message is checked to determine from which source it has come. After determining the type of the received message housekeeping information is added conforming to predefined authorization as per MAMO. Then associated message of the same call is searched in the following step. If the associated message is found both messages are reconciled into one. Otherwise, received message is put into the log for further processing.

When associated message is arrived late compared to its counterpart into the billing site an interruption is occurred to get the already received message from the concerned log. Then do the same as mentioned in the above paragraph.

If the associated message is not received within predefined period a policy decision need to be specified that would be transparent to all the parties. Human intervention is minimized in this framework. Addition and alternation within received message and reconciled message are allowed only as per MAMO.

Reconciled message is then tagged by the time schedule say $T'_1, T'_2, T'_3, ....T'n$. The schedule duration, start time, end time etc. is same with the schedule details of AD switch. Each schedule also has a unique identification number. The objective of tagging scheduling information is to accommodate same call transaction details into both archived files of switch and IN including subscribers' hand set. Tagged reconciled data are putted into storage unit.

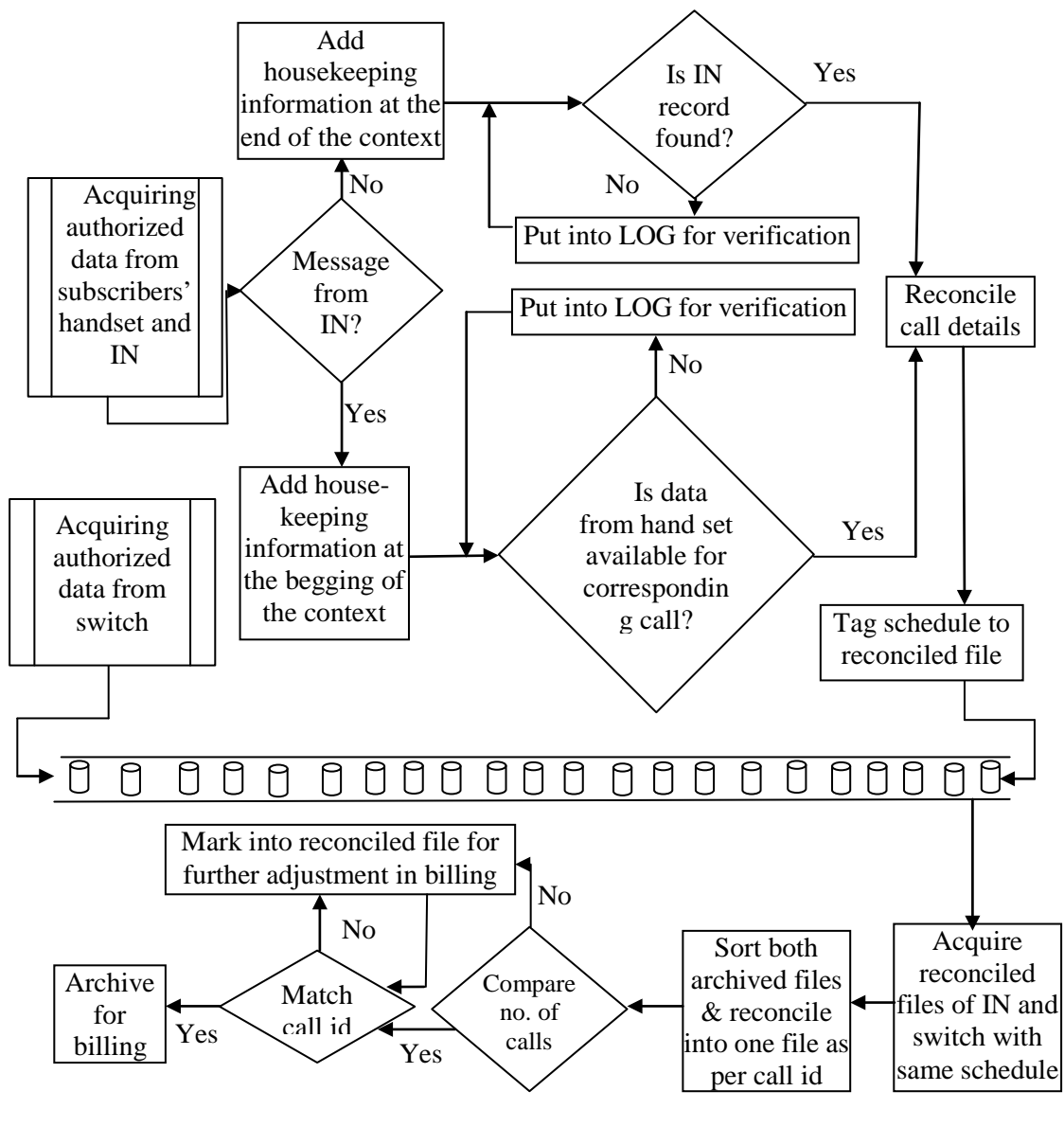

Figure 3. Architecture of Billing System of Extended MAMO

Now $T_i$ and $T'_i$ are fetched by the bottom section of the billing architecture for further processing in asynchronous mode. Both archived files are first sorted then reconciled into one file as per call id. In the next step, check whether numbers of calls are same or not. The outcome of the result is marked into reconciled file. And then call id's are matched against each other for more detailed navigation. Details of unmatched call id's are also marked into reconciled file. Similarly other parameters of call transactions could be contrasted and marked. Here for simplicity only two parameters or attributes are enlisted. Here for simplicity only two parameters are evaluated. In the real scenario a large set of attributes are received for evaluation. Only a small sub set of parameters is jotted down in table 2 for quick read-through. After checking necessary and sufficient parameters bill would be prepared or account balance would be readjusted of prepaid connection.

### 3.3. Implementation

The following characteristics are envisioned on implementation of MAMO to build it professionally acceptable.

- *Imperceptible*: MAMO (rules) is invisible to users and owners. Thus rules need to be mapped by invisible characters that are not shown in the editor.

- *Compatibility*: Compatibility of authorization modes are shown in table 1.

Table 1. Compatibility of Authorization Modes

| AUTHORIZATION MODES | READ ONLY | ADD BEGINNING | ADD END | ADD WITHOUT ALTER | ADD WITH ALTER |
|---|---|---|---|---|---|
| READ ONLY | - | Yes | Yes | No | No |
| ADD BEGINNING | Yes | - | Yes | Yes | No |
| ADD END | Yes | Yes | - | Yes | No |
| ADD WITHOUT ALTER | No | Yes | Yes | - | No |
| ADD WITH ALTER | No | Yes | Yes | No | - |

- *Robustness*: Rules should not be distorted by text processing functionalities. If any rule is distorted by text processing functionalities, then editor is unable to open the segment and consequently a semantic message is thrown.

- *Protection* – Extraneous data are embedded with rules and then encryption is performed. The encrypted rules are inserted into the document.

The Application has been simulated in Java environment. Rich text format is taken as message format and netBeans 6.5 is used as an IDE on Microsoft platform. Different authorizations on multiple messages in a single file are shown in figure 4. The granularity level of authorization is imposed either on a line or a paragraph, or a set of paragraphs.

A message contains different type of information in different parts, referred as sections in the rest of this document. Different access rights may be imposed on different sections by multiple sources. The source may be log information of the handset or calls details of IN. As for example, call duration, source address, destination address, and network availability are acquired from IN are all read only in nature. However, data could be added before all this information maintaining *add before* authorization. Data on signal strength, signal to noise ratio etc are acquired from the handset also read only type located in another section of the message/file. But, data may be added at the end of this information.

Little information needs to be added for housekeeping purposes by the third party and will be located in a third section of the message (see figure 4). All the fields here are assumed for the sake of explaining the concept. Based on the applications and the agreement of services the exact content of the messages would be determined.

Exception Handling: A few typical questions on failure scenarios may occur. A couple of these may be considered like what will happen if handset message reach before the base station message? Or, what happens if the handset message is lost?

A solution to the first problem could be that handset message will be put into the log book for further verification. In the second scenario, a fresh request for the data from handset may be initiated depending upon the terms and conditions for connection. However, as the primary billing parameters are already available; bill may still be generated without reconciliation with handset message.

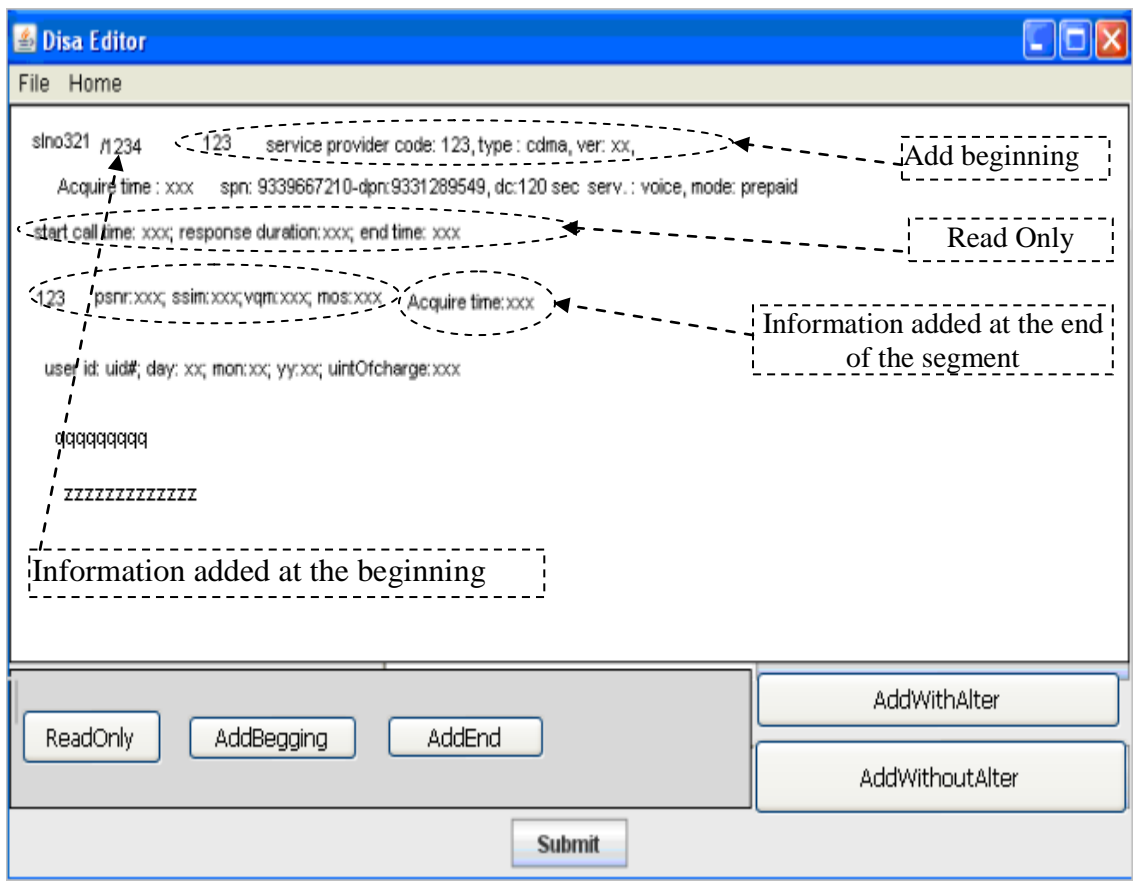

Figure 4. Instance of a sample message for reconciliation

Data is collected from third party billing section for assuring revenue generation, the simulation work is done on number of attributes. Few important attributes are summarized as per chorological order of alphabet in table 2. Few raw data from the communication network to the billing section is also received that are processed as per the requirement. The raw data is marked by the bold letter in the table 2. Data type of all the attributes is of variable length character type with size 255 bytes. Full attribute list is so exhaustive thus avoid here for showing. A few default values of parameters are configured for simulation in the laboratory that may differ in the real life situation. As for example the maximum wait time in a case for getting last partial CDR may be 3 hours that might be changed in other situation.

Table 2. Few Attributes of switch and IN

| ERIC_SWT_T12 (SWITCH) | | SRC_IN_T12 | |
|---|---|---|---|
| Column Name | Column Name | Column Name | Column Name |
| **Filename** | Critoms | **Callidentificationnumber** | Fafindicator |
| Acmchargingindi | Cugindex | **Exchangeidentity** | **Numberofsdpinterrog** |
| Aiurrequested | Cuginterlockcode | **Gsmcallreferencenumber** | Networkid |
| Anmchargingindi | **Callattemptstate** | **Partialoutputrecnum** | Triggertime |
| Acceptablechannelcodings | **Callidentificationnum** | Switchidentity | Redirectioninformation |
| Acccode | Callingsubscriberimeisv | **Recordsequencenumber** | **Dedicatedaccountinformationid** |
| Ageoflocationestimate | Callposition | Mscaddress | **Dedicatedaccountinformationval** |
| Aoccurrencyamountsenttouser | Calledpartymnpinfo | Scfchargingoutputdate | Dedicatedaccvaluebefocall |
| Bcsmtdpdata1servicekey | Calledpartynumnpi | **Outputtype** | Dedicatedaccountvalueaftercall |
| Bcsmtdpdata1gsmscfadd | Calledpartynumton | Chargingunitsaddition | **Accvalueofcallsevents1_acc1** |
| Bcsmtdpdata2servicekey | **Calledpartynum** | Distributed | **Accvalueofcallsevents1_acc2** |
| Bcsmtdpdata2gsmscfadd | Calledsubscriberimei | Freeformatdata | **Accvalueofcallsevents1_acc3** |
| Bcsmtdpdata3servicekey | Calledsubscriberimeisv | Inservicedataeventmodule | **Accvalueofcallsevents2_acc4** |
| Bcsmtdpdata3gsmscfadd | Calledsubscriberimsi | Ssflegid | **Accvalueofcallsevents2_acc5** |
| Bcsmtdpdata4servicekey | Callingpartynumnpi | **Servicefeaturecode** | **Accvalueofcallsevents2_accid** |
| Bcsmtdpdata4gsmscfadd | Callingpartynumton | Single | Accvalueofcallseve2_accvalbit |
| Bcsmtdpdata5servicekey | **Callingpartynum** | Timeforeveinservicedataevemod | Deltavalueofacallevent |
| Bcsmtdpdata5gsmscfadd | Callingsubscriberimei | **Trafficcase** | **Callsetupresultcode** |
| Bcsmtdpdata6servicekey | Callingsubscriberimsi | **Serviceclass** | Subscribernumber_nt |
| Bcsmtdpdata6gsmscfadd | Carrieridentificationcode | **Accountvaluebefore** | Subscribernumber_np |
| Bcsmtdpdata7servicekey | Carrierinfo | **Accountvalueafter** | Subscribernumber |
| Bcsmtdpdata7gsmscfadd | Carrierinfobackward | **Finalchargeofcall** | Originlocinfo_nt |
| Bcsmtdpdata8servicekey | Carrierinfoforward | **Chargedduration** | Originlocinfo_inn |
| Bcsmtdpdata8gsmscfadd | Carrierselecsubstitutioninfo | **Cdrtype** | Originlocinfo_np |

### 3.4. Result

The experiment is simulated with the assistance of computer. Numbers of mobile calls are generated for duration of time in random order. The duration of time is fixed to ten minutes in laboratory experiments.

A random number is assigned at the beginning of every message. The random number is same in both base station message and handset message for every unique call. That random number is matched in both messages for determining whether they belong to the same phone call.

Reconciliation is done on matched messages. The reconciled message is kept in a file for further usage. The average file size and memory usage of the reconciled message is shown in table 3.

Table 3. Average Memory Usage and Average Message Size

| No. of Messages in unit time | Average File Size in KB | Average Memory Usage in KB |
|---|---|---|
| 1000 | 8.56 | 289.85 |
| 5000 | 51.5 | 283.75 |
| 10000 | 105 | 279.6 |
| 15000 | 168 | 272.55 |
| 20000 | 232 | 215.95 |

The file size is varied because it is assumed that different hardware and software configurations of the handset may keep different and variable length data. The reconciliation time is jotted down in figure 5.

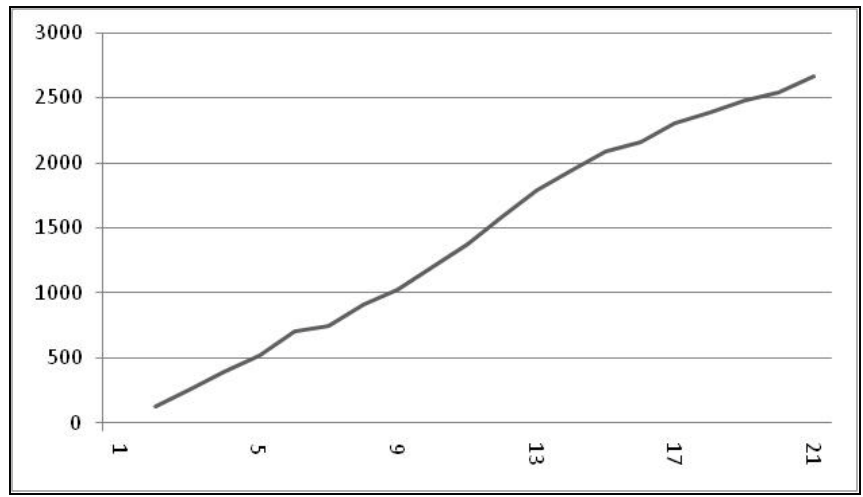

Figure 5. Average reconciliation time in nanosecond vs. number of messages in thousand

Numbers of messages are generated in random order within ten minutes of time duration. Number of messages is implied by the number of phone calls made during that period. Here number of messages implies either the number of messages in the base station or subscribers' handsets for simplicity. Total number of messages is basically double the number of calls held during the stipulated time. The performance of the billing framework grows linearly by the increasing number of phone calls. Total number of phone calls at a time within a cell of the mobile network is constant. The performance of the simulated billing framework is satisfactory for processing that constant number of phone calls at a time. Although in the real life situation input/output cost, traversal cost through network are much higher than our simulation work.

One instance of simulation of revenue assurance is shown in figure 6. Only important parameters are shown for easy reference and simplicity. Data has been collected exhaustively from third party billing organization strictly for academic usage. Computation values of parameters of *Recharge count, total transaction amount, net calculation amount, total transaction balance amount before applying extended MAMO, and then total transaction balance amount after applying extended MAMO* are shown.

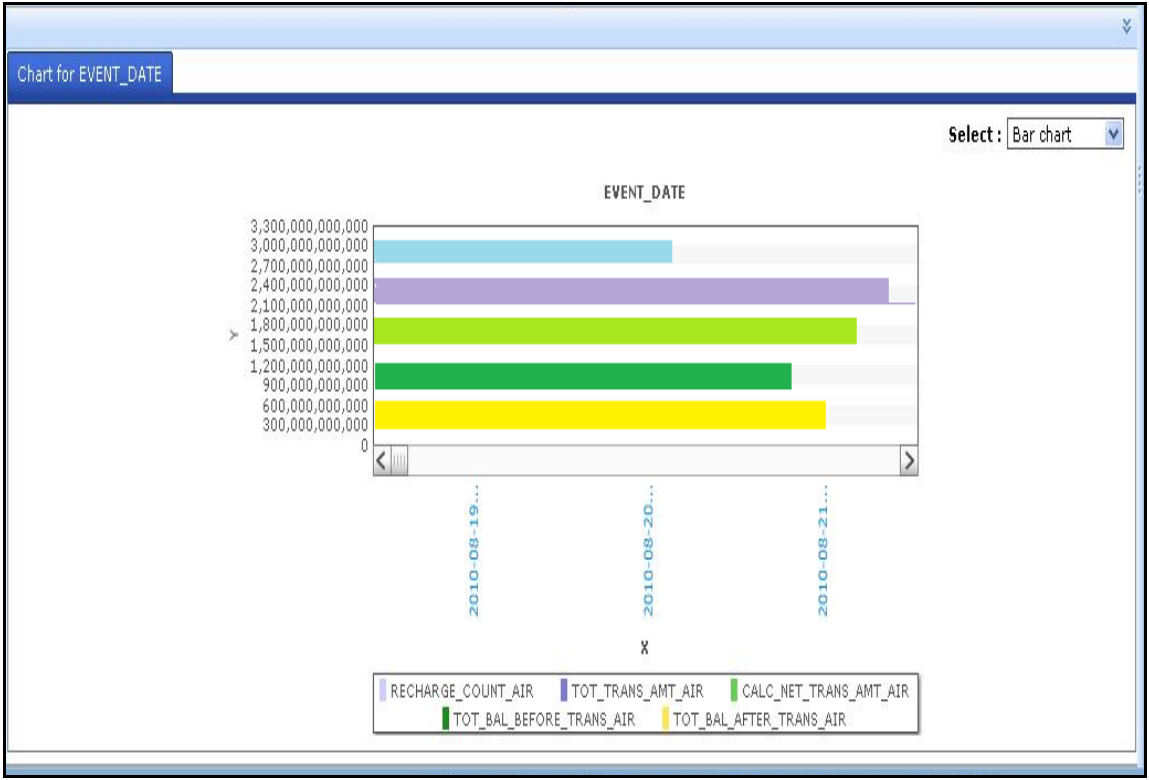

Figure 6. Run time values for different parameters

Data on different service providers from different part of the world is also collected for testing the percentage of increase in revenue. The volume of data differs highly from one another. The output of this simulation varies largely on data volume as it affects other parameters. Or in other words other parameters are dependent on the volume of data. As for example, synchronization problem is much lower between IN and switch on low volume of data. The interdependencies are ignored here, and only percentages of improvements are shown in figure 7.

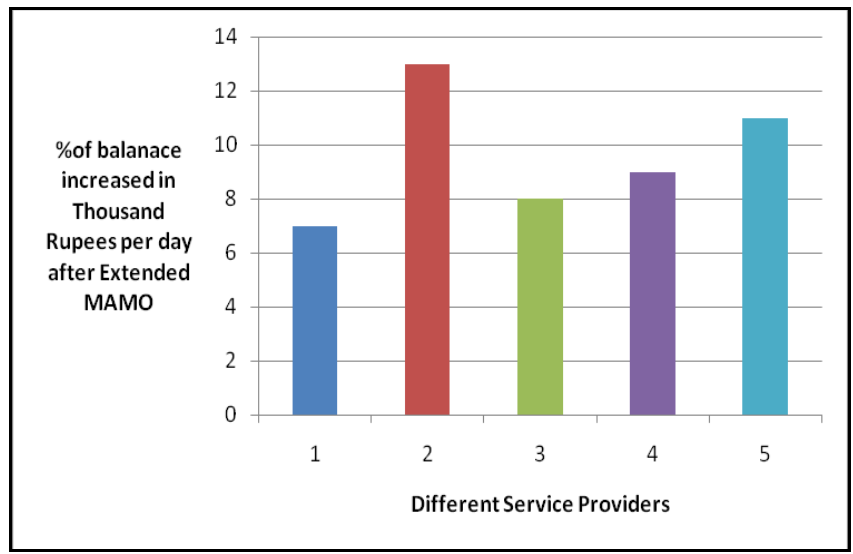

Figure 7. Percentage of increase in Thousand of Rupees on different projects

## 4. CONCLUSIONS

Trustworthiness of the parameters that are provided by the service providers is in question for its early days of usage. Along with, revenue loss is claimed by the service provider due to unexpected behaviour of hardware.

The proper revenue generation of the service provider is assured by the trusted and synchronized billing scheme. The service provider's revenue is scaled up with the accuracy of number of actual transaction occurrence. The trustworthiness on the parameters of subscription is successfully implemented.

**Authors**

Supriya Chakraborty is an Assistant Professor of Department of Computer Application in JIS College of Engineering, West Bengal, India. He received post graduation degree in technology from the National Institute of Technical Teachers Training and Research, Kolkata. Supryo has been involved in research work in the computation and information fields. He has published a few international and national papers and one book imprinted in USA. He has been involved in a number of software and telecommunication industries for training and consultancy. Supriya is always enthusiastic in sports.

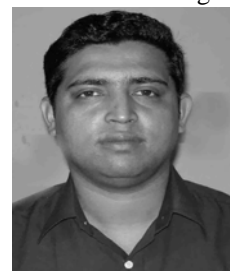

Nabendu Chaki is Head and Associate Professor in the Department of Computer Science & Engineering, University of Calcutta, Kolkata, India. He did his first graduation in Physics and then in Computer Science & Engineering, both from the University of Calcutta. He has completed Ph.D. in 2000 from Jadavpur University, India. Dr. Chaki has authored a couple of text books and more than 70 refereed research papers in Journals and International conferences. His areas of research interests include distributed system, and software engineering. Dr. Chaki has also served as a Research Assistant Professor in the Ph.D. program in Software Engineering in U.S. Naval Postgraduate School, Monterey, CA. He is a visiting faculty member for many Universities including the University of Ca'Foscari, Venice,

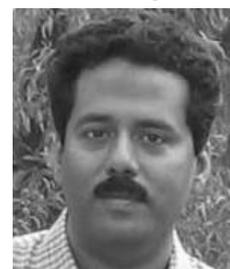

Italy. Dr. Chaki is a Knowledge Area Editor in Mathematical Foundation for the SWEBOK project of the IEEE Computer Society. Besides being in the editorial board for several International Journals, he has also served in the committees of more than 50 international conferences.